# A new approach to the step-drawdown test

*Dr. Gianpietro Summa*
*Ph.D. in Environmental and Applied Geology*
*Monticchio Bagni*
*85028 Rionero in Vulture (Potenza), Italy*
*e-mail: info@gianpietrosumma.it*

**Abstract**
In this paper a new approach to perform step-drawdown tests in presented. Step-drawdown tests known so far are performed strictly keeping the value of the pumping rates constant through all the steps of the test. Current technology allows us to let the submerged electric pumps work at a specific revolution per minute (rpm) and allows us to suitably modify the rotation velocity at every step. Our approach is based on the idea of keeping the value of rpm fixed at every step of the test, instead of keeping constant the value of the discharge. Our technique has been experimentally applied to a well and a description of the operations and results are thoroughly presented. Our approach, in this peculiar case, has made possible to understand how actually the discharge $Q$ varies in function of the drawdown $s_w$. It has also made possible to monitor the approaching of the equilibrium between $Q$ and $s_w$, using both the variation of $Q$ and that of $s_w$ with time. Moreover, we have seen that for the well studied in this paper the ratio $Q/s_w$ remains almost constant within each step.

**Introduction**
Step-drawdown tests are nowadays quite popular: they are the most frequently performed tests in the case of single well (Kawecki 1995). There are various reasons why they are performed: in the case of exploration wells, they allow to determine the proper discharge rate for the subsequent aquifer test; in the case of exploitation wells they allow, among other things, to understand the behavior of the well during the pumping, to determine the optimum production capacity and to analyze the well performance over time (Boonstra and Kselik 2001).
Step-drawdown tests have been introduced by Jacob (1947) in order to study how the discharge $Q$ affects the drawdown $s_w$ of a well drilled in confined aquifers. He proposed the following equation:

$$s_w = BQ + CQ^2 \qquad (1)$$

where $B$ is a numerical parameter which takes into account the linear head losses of the aquifer, while $C$ is the analogous parameter for the non-linear (quadratic) head losses, mainly imputable to the construction characteristics of the well (Driscoll 1986). Jacob (1947), deriving his equation, made the analogy between the drawdown $s_w$ of a well and the voltage drop $\Delta V$ through a resistor, and between the discharge $Q$ and the electric current $I$.
Under steady-state conditions, for confined aquifers and applying Dupuit's assumption, the coefficient $B$ in equation (1) is constant with time and it can be directly derived from Dupuit's formula.

Instead, under unsteady-state (transient) conditions, the coefficient *B* in equation (1) is time-dependent and, for confined aquifers, it may be represented through the Cooper-Jacob approximation of the Theis solution for an infinite uniform aquifer (van Tonder et al. 2001). However, if the aquifer is infinite and not recharged, the radius *R* of action of the well scales as $\sqrt{t}$ with time, and its time-derivative *dR/dt* scales as $1/\sqrt{t}$. If *t* is large, *dR/dt* is close to zero, namely *R* varies very slowly and it seems as if a steady-state has been achieved (de Marsily 1986). The relations for confined aquifers are also applicable to unconfined aquifers as long as the drawdown is small in comparison with the aquifer thickness (Driscoll 1986). The scientific literature on this topic is vast and variegated (see, for example, Krusemar and de Ridder (1994) and Driscoll (1986) and the references quoted therein).

The step-drawdown test is the first step after a sequence of practical operations carried out for the construction of a new well. However, some authors have used the step-drawdown test to evaluate drawdowns in unconfined, heterogeneous and anisotropic aquifers with good results (Helweg 1994).

In this paper we focus the attention mainly on the way in which the tests are performed, rather than on their interpretation, and the final goal is to provide the community with tools useful to draw a better characteristic curve of a well.

**Preliminary operations**
After the well is drilled, preferably with percussion method, pumping tests are carried out: usually they are step-drawdown tests. In order to complete the well construction we need to know the maximum discharge obtainable, namely the maximum discharge reachable by emptying the well with an electric pump placed at the bottom.

Sometimes, due to the high transmissivity of some aquifers and/or to the low power of the electric pumps placed into the well, the above test condition can not be fulfilled. Nevertheless, it is possible to gain anyway the necessary information for the screen transmitting capacity.

If the borehole walls are stable, then it is possible to proceed with the pumping test without protections to the electric pump; instead, if the borehole walls are unstable, the pump can be shielded with a tube within which the pump is suitably placed. In case in which the walls of the borehole are so weak to collapse, it is necessary to install a temporary well screen (for example: a Johnson screen) with the greatest slot openings available and a rough gravel pack in the space between the well screen and the borehole. Sometimes it is also necessary to place some gravel coarseness at the bottom of the well screen. At this point an electric pump capable of emptying completely the well has to be placed inside it. If there were no such electric pumps, then an electric pump able to create the greatest possible drawdown should be used.

In the ideal conditions in which the well is uncasing and it has been emptied by the electric pump, we start to obtain the discharge due exclusively to the aquifer. With such value of the discharge we should be able to choice the right dimension of the well screen to be installed definitely in the well. Usually, such choice is made taking into account the whole filtering surface, that typically extends itself over the entire saturated portion of the aquifer.

In order to perform good step-drawdown tests we need to determine the dimensions the well screen in such a way that the water intake velocity is less than 0.03 m/s (Driscoll 1986), even under the greatest possible pumping rate (and thus under the greatest possible drawdown).

For this reason, it is usually worth to determine the dimensions of the well screen taking into account only a short portion (2-4 meters, to be evaluated case by case) placed immediately above the pump intake. In this way we reduce to the minimum the risk of a turbulent flow inside the well. Also for its action against the sand, one should choice the Johnson screen (Driscoll 1986).

It is preferable to extend the tube below the well screen with a short piece of tube with closed bottom. In this way we are able to install the electric pump with the water intake slightly below the well screen; moreover, sediments accumulated in such tube extension allows us to evaluate the amount of detritus brought by water and finally, makes the cleaning operations of the well easier.

Soon after the well screen has been installed, the construction of the well can be completed with the insertion between the tube and the borehole of a suitable gravel pack along the whole length of the screen. The non-filtering portion of the interstice between the tube and the borehole must be suitably sealed.

Quite often it is used to put cement grout in the interstice piece by piece in order to avoid the crushing of the tubes. Probably, it would be better to seal the interstice alternating cement grout and bentonite.

In this phase, the electric pump can be placed near the bottom of the well and can be turned on to provide the maximum discharge for the development procedures.

It would be better to use an electric pump without non-return valve. In this way, during the stops of the pump, the naturally generated inverse flux can destroy the sand bridges created during the maximum discharge phase (Driscoll 1986). Only when pumped water is clean then one should shut off the pump and wait the least time needed to the well to restore its static level. Hence, the pump should be turned on and off many times until the pumped water is clean just from the beginning of the pumping.

**Description of the operative method**

In order to perform the step-drawdown test the electric pump must be driven by a frequency drive: in this way we are able to change the rotation velocity in a smooth manner. The discharge flow has to be measured by an electromagnetic flow meter (Fig. 1). The drawdown must be measured with a electric sounder and with a level probe placed immediately above the electric pump, at a fixed depth.

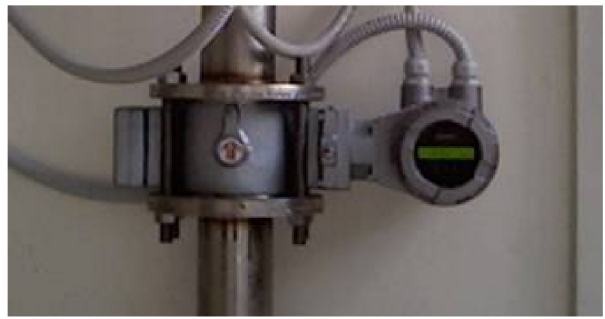

**Fig. 1: Electromagnetic flow meter. Accuracy $1 \times 10^{-5}$ m$^3$/s**

Here the use of an electromagnetic flow meter is essential since it allow us to measure the flow instantaneously, while the electric sounder, together with a level probe inside the well, provides us with safe, continuous measurements of the drawdown.

Thus, it is possible to measure the instantaneous discharge and compare it with the relative instantaneous level measure.

Now, it is possible to plan the execution of the step-drawdown test and choice the more suitable number of steps: according to Krusemar and de Ridder (1994) the minimum should be 3.

It is also important for the test to check the maximum and minimum rotation velocity of the electric pump. For the minimum rotation velocity we can turn the pump on and gradually increase the rotation velocity till when the pumped water does not reach the height of the surface. The velocity satisfying such condition is the minimum velocity. Instead, the maximum velocity is equal to the maximum rotation rate of the pump, if the electric pump is not able to empty the well. Otherwise, it can determined turning the electric pump on at its maximum velocity and then gradually reducing the rotation rate till when the flow oscillations measured by the electromagnetic flow meter will be damped: namely, the maximum rotation velocity will be the velocity corresponding to a stable and continuous discharge.

After the determination of the maximum and minimum velocities, it is important to wait a suitable interval of time needed to the well to restore its static level. Only after such time it is possible to start the step-drawdown test.

Since we know the range of the rotation velocity, now we can choice the number of the steps and the relative velocities to proceed with the test.

Suppose, for example, that we have obtained a minimum rotation velocity of 1800 rpm and a maximum rotation velocity of 2800 rpm. Since we want to perform a three-steps test, we could choice the following values: $1^{st}$ step at 1900 rpm; $2^{nd}$ step 2100 rpm e $3^{rd}$ step at 2700 rpm. Obviously, the more the steps are, the better the drawing of the characteristic curve will be.

With the above approach, it is possible to draw the first part of the characteristic curve with higher precision since we are able to increase the discharge very gradually with the frequency drive.

After the decision on the total number of the steps, we start the test form the first one, namely that with the smallest rotation velocity.

Now, in our tests what is constant is no more the discharge, but the rotation velocity of the pump. It is important to collect the discharge and the drawdown values after a fixed interval of time, e.g. once per minute.

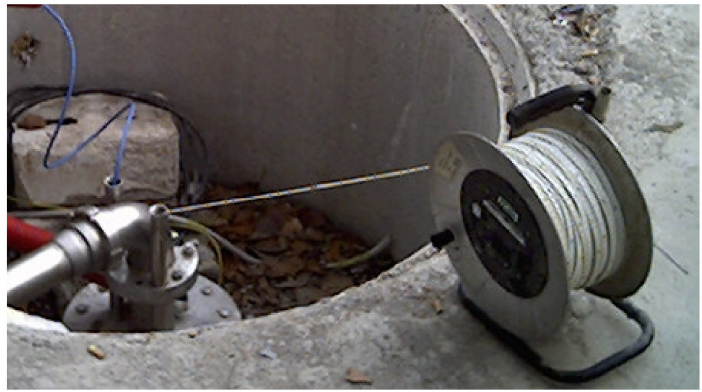

**Fig. 2: Measurements of water level with an electric sounder**

If a level probe has been placed inside the well, then the electric sounder is used as a check instrument. The measurement intervals with the electric sounder (Fig. 2) can be chosen to be, for example, once per minute for the first ten minutes, once every five minutes for the first half an hour, then once every ten minutes for the first hour; from the first hour on, every thirty minutes (Table 1).

| time | Step 1 rpm 2110 | | Step 2 rpm 2210 | | Step 3 rpm 2320 | | Step 4 rpm 2410 | |
|---|---|---|---|---|---|---|---|---|
| | $Q_1$ x $10^{-3}$ | $s_w1$ | $Q_2$ x $10^{-3}$ | $s_w2$ | $Q_3$ x $10^{-3}$ | $s_w3$ | $Q_4$ x $10^{-3}$ | $s_w4$ |
| (min) | (m³/s) | (m) | (m³/s) | (m) | (m³/s) | (m) | (m³/s) | (m) |
| 0 | 0.00 | 0.00 | 4.60 | 36.02 | 5.28 | 44.24 | 5.90 | 53.80 |
| 1 | 8.20 | 11.63 | 6.20 | 38.00 | 6.91 | 46.35 | 7.12 | 55.19 |
| 2 | 8.00 | 14.93 | 6.00 | 38.90 | 6.83 | 47.30 | 7.07 | 55.78 |
| 3 | 7.60 | 17.20 | 6.04 | 39.70 | 6.73 | 47.94 | 7.00 | 56.20 |
| 4 | 7.30 | 19.00 | 5.95 | 39.96 | 6.67 | 48.43 | 6.95 | 56.52 |
| 5 | 7.10 | 20.40 | 5.80 | 40.30 | 6.59 | 48.81 | 6.93 | 56.80 |
| 6 | 6.85 | 21.59 | 5.84 | 40.63 | 6.57 | 49.14 | 6.90 | 57.03 |
| 7 | 6.65 | 22.50 | 5.80 | 40.86 | 6.54 | 49.41 | 6.92 | 57.24 |
| 8 | 6.50 | 23.30 | 5.70 | 41.06 | 6.47 | 49.64 | 6.92 | 57.43 |
| 9 | 6.27 | 24.00 | 5.75 | 41.26 | 6.42 | 49.85 | 6.91 | 57.60 |
| 10 | 6.26 | 24.61 | 5.70 | 41.41 | 6.40 | 50.04 | 6.87 | 57.74 |
| 15 | 6.00 | 26.83 | 5.67 | 42.03 | 6.40 | 50.74 | 6.78 | 58.33 |
| 20 | 5.70 | 28.26 | 5.56 | 42.45 | 6.24 | 51.14 | 6.71 | 58.76 |
| 25 | 5.50 | 29.29 | 5.57 | 42.75 | 6.24 | 51.45 | 6.72 | 59.09 |
| 30 | 5.40 | 30.11 | 5.48 | 42.98 | 6.25 | 51.75 | 6.64 | 59.33 |
| 40 | 5.21 | 31.27 | 5.40 | 43.37 | 6.07 | 52.18 | 6.54 | 59.71 |
| 50 | 5.10 | 32.11 | 5.35 | 43.61 | 6.07 | 52.49 | 6.54 | 59.98 |
| 60 | 5.00 | 32.75 | 5.33 | 43.83 | 6.05 | 52.72 | 6.44 | 60.18 |
| 90 | 4.90 | 34.03 | 5.28 | 44.24 | 6.02 | 53.22 | 6.40 | 60.50 |
| 120 | 4.75 | 34.71 | 5.28 | 44.24 | 5.96 | 53.50 | 6.40 | 60.50 |
| 150 | 4.68 | 35.21 | 5.28 | 44.24 | 5.90 | 53.80 | 6.40 | 60.50 |
| 180 | 4.66 | 35.55 | | | 5.90 | 53.80 | | |
| 210 | 4.62 | 35.82 | | | 5.90 | 53.80 | | |
| 240 | 4.60 | 36.02 | | | | | | |
| 270 | 4.60 | 36.02 | | | | | | |
| 300 | 4.60 | 36.02 | | | | | | |

**Table 1: Step-drawdown test carried out on November 8, 2003. The red figures refer to the values of the discharge and of the drawdown of the steps (each of which lasts 60 min) measured soon after the beginning of the stabilization. The increase in rpm between the steps requires few seconds to be achieved and, for this reason, for each step subsequent to the first one the duration time is counted as in column 1 and the values of $Q$ and $s_w$ at each step time $t=0$ are those relative to the stabilization point of the previous step**

Soon after the pump is turned on, the discharge begins a decreasing trend in time, while the drawdown starts to increase (Fig. 3). After a suitable interval of time, during which a pseudo steady-state or steady-state flow is approached, both the discharge and the drawdown reach a stabilization point. After the stabilization, it is important to check that the values of the discharge and those of the drawdown do not change over an interval of time that must be equal for every step of the whole test: this time just defines the duration of the steps, and it can be chosen to be, for example, of 60 minutes from the beginning of the stabilization. All this guarantees that the data gathered in every step are homogeneous and can be safely and significantly compared.

In our approach the time duration of the whole test is not taken into account as a useful parameter for the subsequent analysis, since $Q$ in constant only over a portion of the entire duration of a step: what is important is the time duration of each single step and the behavior of $Q$ and $s_w$ inside each single step. Moreover, the transition from a previous step to the following one needs a few seconds interval for the increase in the rotation velocity of the pump (which can't be made instantaneous) and such technical times are not counted. Therefore the time counting is reset to zero at the beginning of every step, provided that, for the steps subsequent to the first one, the values of $Q$ and $s_w$ at the step time $t=0$ are those relative to the stabilization point of the previous step (see Table 1).

After having collected the data of the first step, starting from the beginning till the stabilization and for the following 60 minutes, we suddenly increase the rotation velocity of the pump to the value chosen for the execution of the second step: now, we repeat all the operations carried out for the first step, and so on for all the steps planned for our step-drawdown test.

Once the data acquisition is completed on all the steps, we have on hand as many pairs of $Q$ and $s_w$ as the performed steps and, among other things, the characteristic curve of the well can be drawn.

One of the main advantage of our approach is that we do not have to proceed with manoeuvres which may perturb the system as, for example, those needed to fix the discharge during the execution of a classical step-test at constant rate (Castany 1982).

Moreover, in our approach we can draw the curves $Q$-$t$ and $s_w$-$t$ for every step: since these trends are free from disturbance given by every possible regulation manoeuvres, they reveal themselves to be useful also in the study of the well-aquifer system.

**Results from an actual experiment**

Here we present the results of an actual step-drawdown test carried out on a well that taps a confined and heterogeneous aquifer in the Southern Italy. The well, already described in Piscopo and Summa (2007), has been drilled following the operative method described in the previous paragraph, unless for the length of the well screen that, for financial reasons, has been reduced to nearly 1/3 of the aquifer thickness.

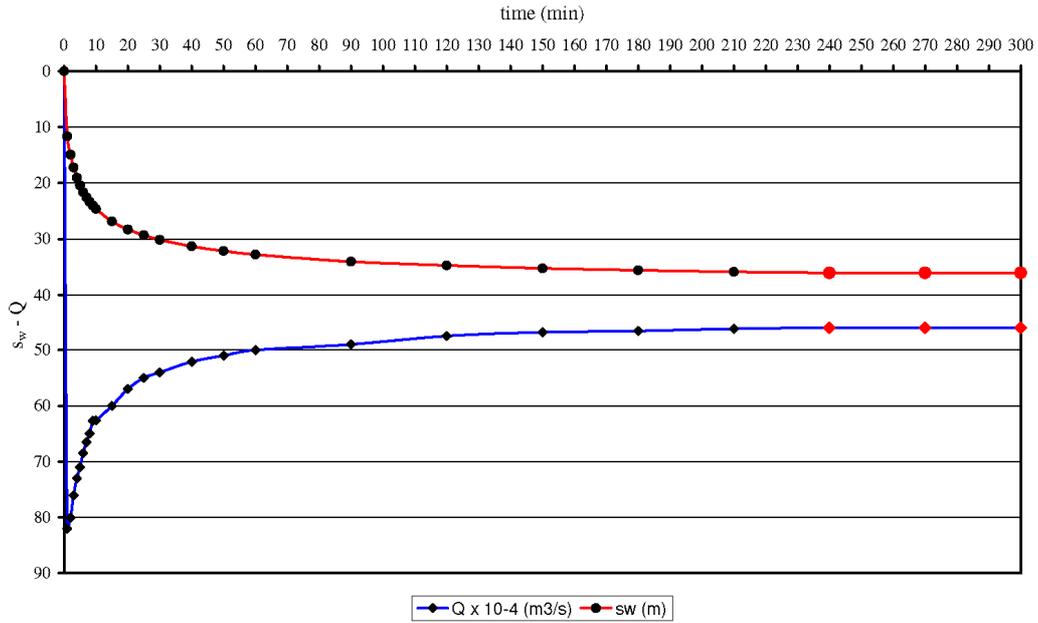

**Fig. 3: Time-drawdown and time-discharge plots for the first step with rotation velocity of 2110 rpm. Red dots are used after the stabilization**

We have performed 4 steps. The results are listed in Table 1. For every step in discharge we have drawn the *Q-t* and $s_w$ -t curves, as shown in Fig. 3. As one can easily note in Fig. 3, for fixed rotation velocity the discharge decreases and the drawdown increases quite rapidly at the beginning of the test, while, starting from nearly 240 minutes after the beginning, these values stabilize and from now on we can count the duration of the step in the step-drawdown test, previously fixed in 60 minutes. It is interesting to plot $s_w$ against *Q*; the result is showed in Fig. 4.

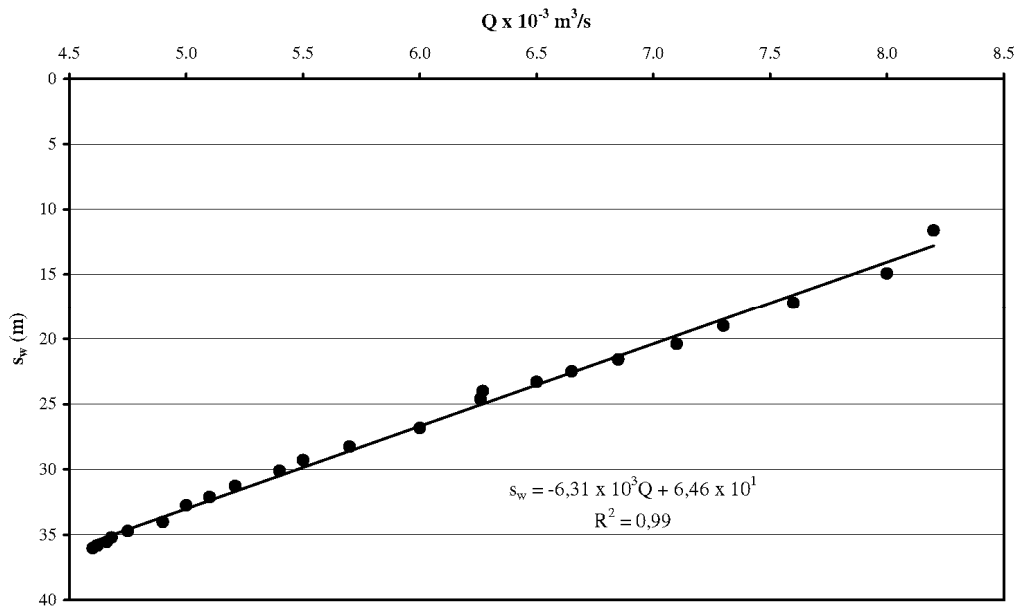

**Fig. 4: The drawdown $s_w$ as a function of *Q* for the first step (2110 rpm) before the stabilization**

After 60 minutes the *Q* and $s_w$ parameters of the first step is considered to be stabilized; they are recorded and then we proceed with the test. We suddenly increasing the rotation velocity to the value previously planned for the 2$^{nd}$ step. The measured values for $Q$ and $s_w$ in this case are shown in Fig. 5.

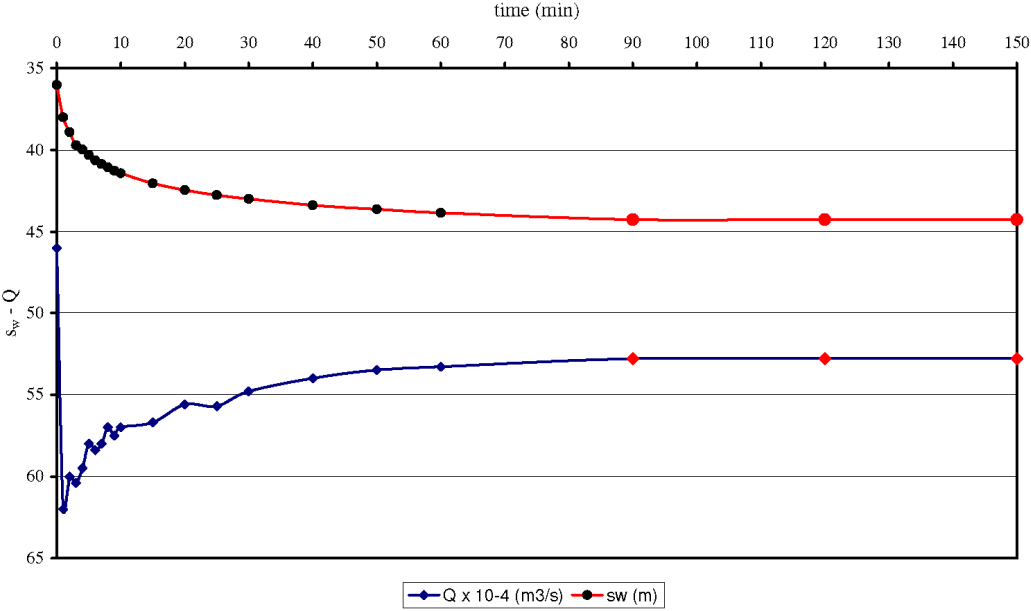

**Fig. 5: Time-drawdown and time-discharge for the second step at 2210 rpm. Red dots are used after the stabilization**

As for the 1$^{st}$ step, we wait until the new stabilization point is reached (nearly after 90 min) and then we record the values of $Q$ and $s_w$ for the following 60 minutes. In the same way, we proceed with all the following planned steps. In figures Fig. 6 and Fig. 7 the results for the 3$^{rd}$ and 4$^{th}$ steps are shown. In Table 2 we lists the pairs $Q$ and $s_w$ recorded after the stabilization for all the four steps. Now, the characteristic curve of the well can be drawn (Fig. 8).

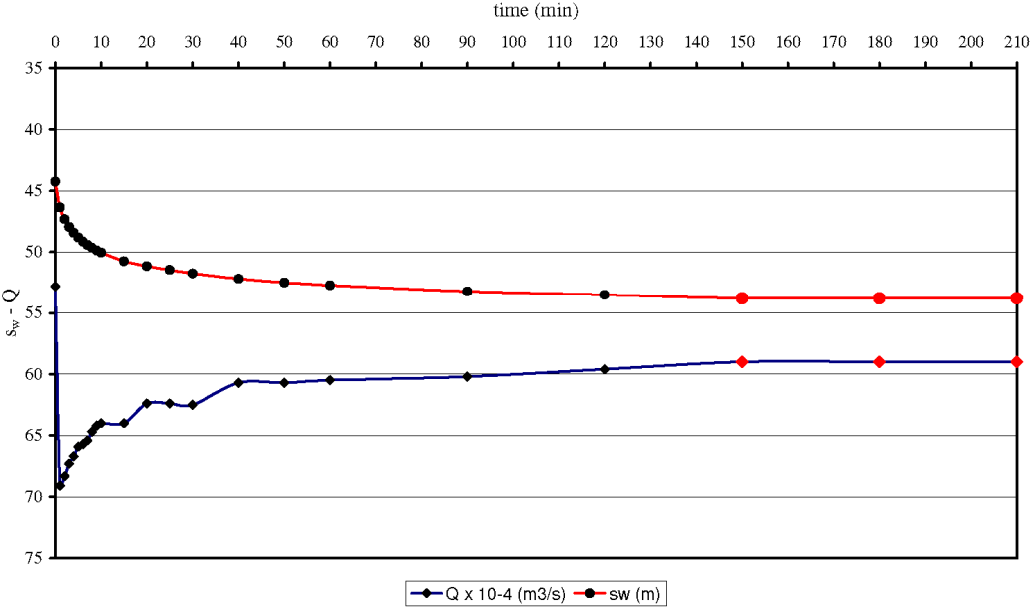

**Fig. 6: Time-drawdown and time-discharge for the third step at 2320 rpm. Red dots are used after the stabilization**

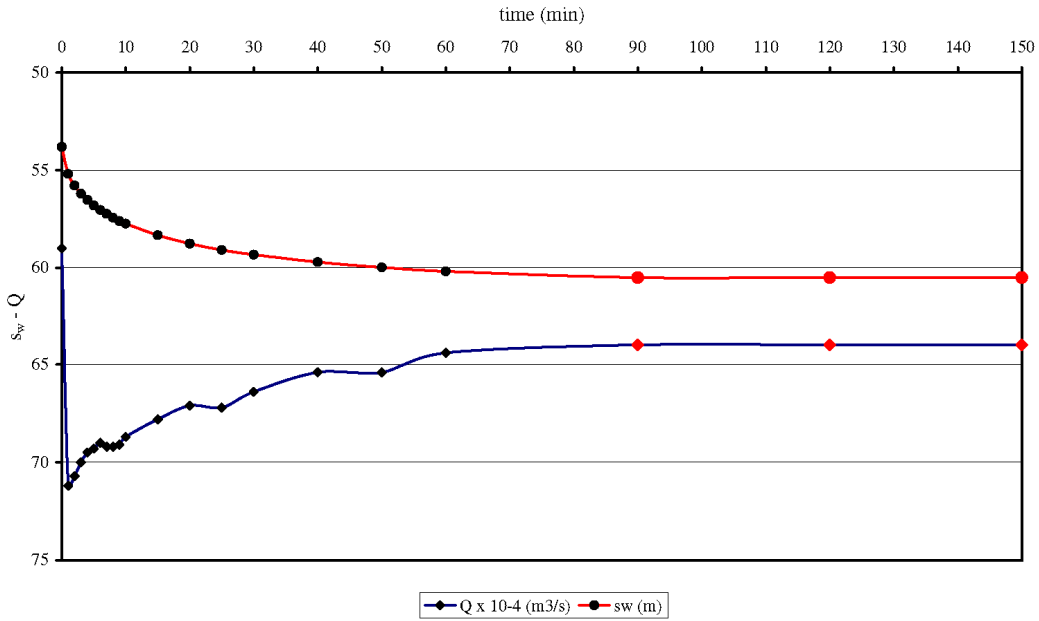

**Fig. 7: Time-drawdown and time-discharge for the fourth step at 2410 rpm. Red dots are used after the stabilization**

Once the step-drawdown test is performed, it is possible to complete the construction of the well and let it to become operational. As a matter of fact, all the information needed for a proper use of the well, namely the stratigraphy, the well design and the characteristic curve, are now available.

|                     | Q<br>x $10^{-3}$ (m$^3$/s) | $s_w$<br>(m) |
|---------------------|---------------------------|--------------|
| step 1 - rpm 2110   | 4.60                      | 36.02        |
| step 2 - rpm 2210   | 5.28                      | 44.24        |
| step 3 - rpm 2320   | 5.90                      | 53.80        |
| step 4 - rpm 2410   | 6.40                      | 60.50        |

**Table 2: Pairs $Q$ and $s_w$ recorded after the stabilization for the four steps of the test**

It is important to stress that the characteristic curve of a well holds its validity only if it is provided with the execution date of the step-drawdown test, therefore *B* and *C* are parameters useful to the proper representation of the functional relationship between *Q* and $s_w$. As a matter of fact, the results of a step-drawdown test and the corresponding characteristic curve strongly depends on the period of the year in which they are carried out, on the specific year and on some possible anthropic interferences.

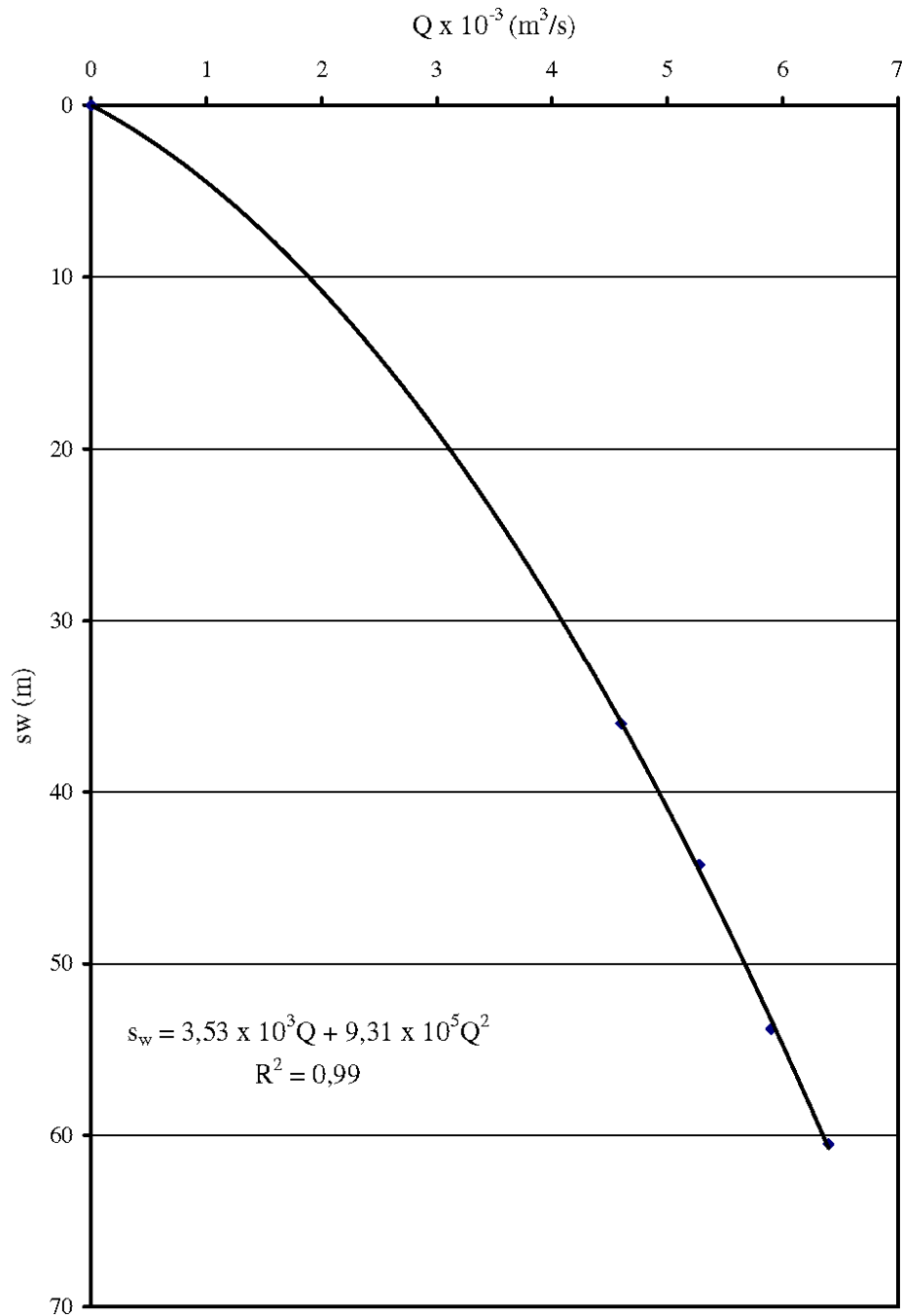

**Fig. 8: Characteristic curve of the well presented in this section (November 8, 2003)**

If the preliminary pumping tests (see the Preliminary operations Section) were carried out in the same way as the final step-drawdown test, then it would be possible to significantly compare the results in order to evaluate the changes occurred to the well due to the installation of the well screen and to the completion procedures. In such a way the final step-drawdown test would become a sort of test of the well.

**Discussion**

The characteristic curve, derived as before, provides us with useful information on the hydraulic behavior of the well and gives also information on its fair exploitation.

The step-drawdown test carried out in the previous section shows, among other things, how for every steps executed at fixed rotation rate it is not possible to have changes in the drawdown without having corresponding changes in the discharge. As a matter of fact, the characteristic curve is the graphical representation of such functional relation.

If we remind the analogy proposed by Jacob (1947) between $Q$ and the electric current $I$, and between $s_w$ and the voltage drop $\Delta V$, then the curve in the plane ($Q$, $s_w$) is comparable to the characteristic curve of an elementary electrical device in the plane ($\Delta V$, $I$), and thus it can be seen as representing the well itself.

Some comparisons between the classical test at constant discharge and our approach follow in order. As a matter of fact, during a classical test at constant discharge $Q$ it is possible to record only the drawdown $s_w$ and there is a very high probability that the measurement of the constant discharge $Q$ results to be imprecise. Quite often the measurement of the discharge $Q$ is made with indirect methods as the turbine water meter or the Woltmann water meter. All these measurement devices require a measurement time which is long if compared with the time needed for the measurement of the drawdown $s_w$. Thus, the measured discharge is only but a mean value over the measurement time.

On the other side, the direct and instantaneous measurement provided by the electromagnetic flow meter in our approach, other than being more precise, allows a straight comparison with the corresponding drawdown values.

Moreover, the results of the step drawdown tests carried out with our method keep the property of being readable, analyzable and exploitable in the frame of the classical approaches developed in the past years, starting form Jacob (1947). Besides, our approach provides also additional information about the discharge variations with time for each steps: such data can be used for further analyses.

For example, it is interesting to show, together with the characteristic curve on the plane ($Q$, $s_w$), the trend of $s_w$ as a function of $Q$ for each step before the stabilization: in Fig. 9 the continuous line is the characteristic curve, while the dots with different color and shape represent the dependence of $s_w$ on $Q$ during the approach to the stabilization (top-down) in each step; the stabilization points are the intersections between the characteristic curve and the lines which interpolate each sequence of colored dots.

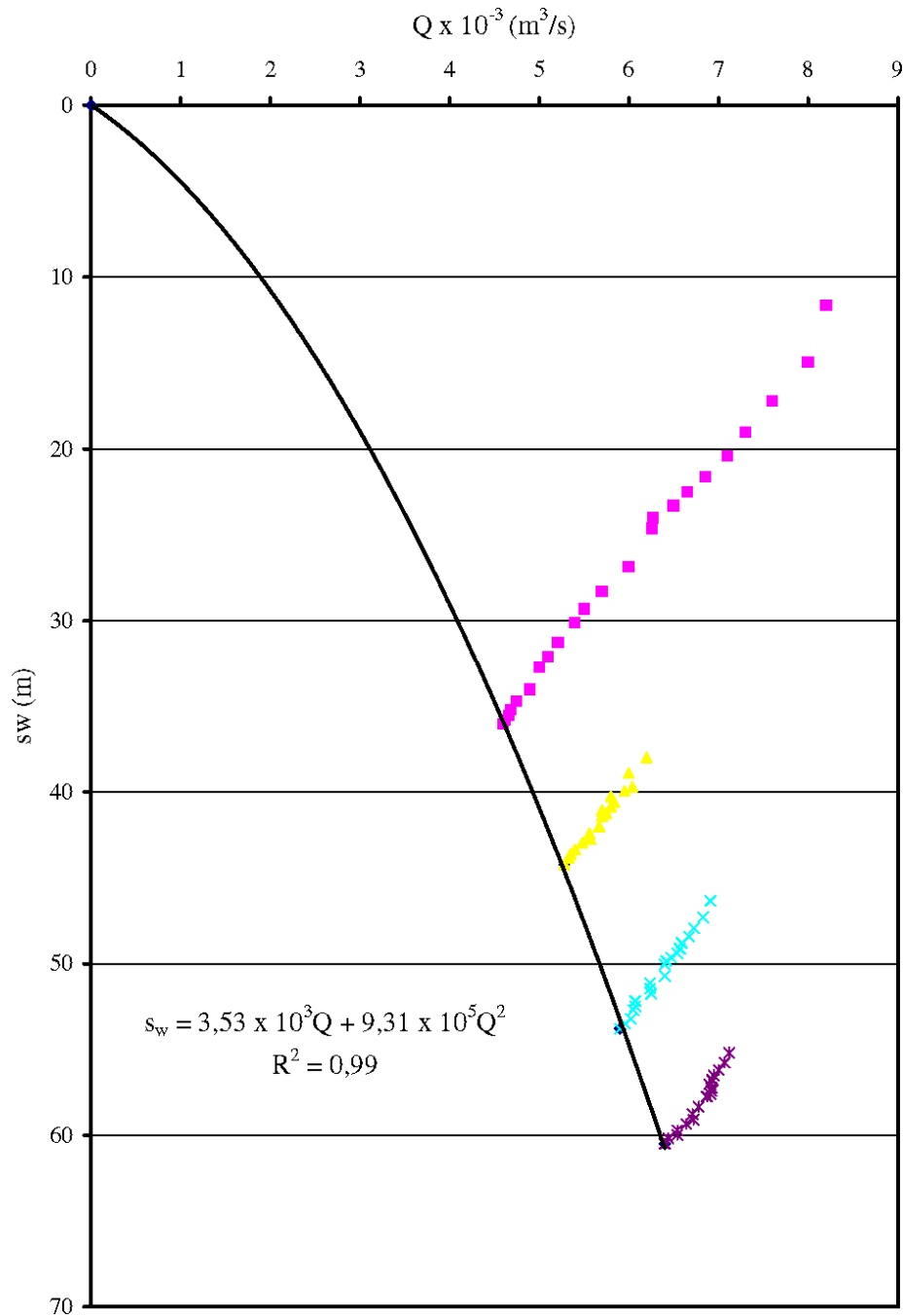

**Fig. 9:** The continuous black line is the characteristic curve, while the dots with different color and shape represent the dependence of $s_w$ on $Q$ during the approach to the stabilization (top-down) for each step of the test; the stabilization points are the intersections between the characteristic curve and the lines which interpolate each sequence of colored dots

We want to stress here that the most valuable information on the well is provided to the hydrogeologist by the characteristic curve or, better, by the characteristic curves: they describe the hydraulic behavior of the system well-aquifer under various circumstances, also they account for aquifer perturbations, both natural and artificial (man made).

As a matter of fact, it would be better to have at least two characteristic curves of the well, one obtained during the drought period of the aquifer, and the other obtained during the recharge period.

The characteristic curves are indeed precious instruments for the judgment of the hydrogeologist which has to carry out a fair exploitation of the well: for example, they allow the hydrogeologist to evaluate the suitable discharge or the suitable drawdown to be applied for a steady use of the well, or they allow to establish what is the maximum discharge safely extractable for few hours during a given period of the year. As a matter of fact, given a precise request on the discharge of the well by the customer, the characteristic curve allows to choice the right pumping parameters to satisfy the demand. For example, if one needs $5 \times 10^{-3}$ m$^3$/s from the well studied above, then we find on the characteristic curve the drawdown corresponding to a discharge of $5 \times 10^{-3}$ m$^3$/s, and we take such a value as parameter for the steady use of the well, as explained in Piscopo and Summa (2007). If the characteristic curve is a curve obtained during a drought period of the aquifer, then the required discharge will be with high probability always available.

**Conclusions**

Here we shortly summarize the main results of our paper. The new approach to the step-drawdown tests presented in this paper allows:

- to greatly reduce the perturbations in the water flow which usually are generated during the classical tests at constant discharge;
- to instantaneously record the values of the drawdown and those of the discharge with fine time resolution, from the beginning of the step to the stabilization;
- to perform actual tests of the same duration for each step, once the stabilization is reached;
- to rapidly draw a precise characteristic curve of the well, and thus to give an hydraulic characterization of the system well-aquifer;
- to perform also the usual analyses cited in the literature;
- to carry out new analyses on the variation of the discharge $Q$ with time $t$, in each step.

Moreover, we think that our approach poses new interesting theoretical questions: for example, during the drilling operation of the borehole, when a discharge test is carried out at open borehole with the aim of filter dimensioning, if the dimensioning procedure is performed correctly, what is the new meaning of the parameter $C$, introduced by Jacob (1947)?

Besides, a comparison between the results of the final discharge test and the results of the test performed at open borehole could be see also as a direct tests of the well itself.

Finally, the characteristic curve, other than being a tool useful to decide a suitable and fair exploitation of the well, unequivocally shows that it is not possible to have a drawdown without a corresponding variation of the discharge.

**Acknowledgements**

The author would like to thank Futurella Company, in the person of Mr. Nicola Del Negro, for the opportunity to conduct the field work. The author is grateful to Prof. Germano D'Abramo for his close friendship. The author is also grateful to Prof. Vincenzo Piscopo and Prof. Pantaleone de Vita, and to Dr. Salvatore Grimaldi for a critical reading of the manuscript. A special thanks goes to Dr. Assunta Tataranni, support of my life.